\documentstyle[aps,prl,twocolumn,graphics,epsfig,floats,array]{revtex}
\newcommand{\be}{\begin{eqnarray}}
\newcommand{\ee}{\end{eqnarray}}

\def\l{\langle}
\def\r{\rangle}

\begin{document}

\title{Robustness of Multi-Party Entanglement}

\author{Christoph Simon$^{1,}$\thanks{email:christoph.simon@qubit.org} and Julia Kempe$^{2,3,}$\thanks{email:kempe@math.berkeley.edu}}
\address {$^1$ Centre for Quantum Computation, University of Oxford, Parks Road, Oxford OX1 3PU, United Kingdom\\  $^2$ Departments
of Mathematics and Chemistry, University of California, Berkeley\\ $^3$ CNRS-LRI, UMR 8623, Universit\'e de
Paris-Sud, 91405 Orsay, France}
\date{\today}
\maketitle

\begin{abstract}
How common is large-scale entanglement in nature? As a first step towards addressing this question, we study the
robustness of multi-party entanglement under local decoherence, modeled by partially depolarizing channels acting
independently on each subsystem. Surprisingly, we find that $n$-qubit GHZ entanglement can stand more than 55 \%
local depolarization in the limit $n
\rightarrow \infty$, and that GHZ states are more robust than other generic states of 3 and 4 qubits. We also study
spin-squeezed states in the limit $n
\rightarrow \infty$ and find that they too can stand considerable local depolarization.

\end{abstract}

\bigskip

Entanglement is certainly one of the most dramatically non-classical features of quantum physics. There are states
of composite quantum systems that cannot be decomposed into probabilistic combinations of product states: there is
no way of writing the density matrix of such a state $\rho$ in the form $\rho=\sum
\limits_i p_i \sigma_1^i \otimes \sigma_2^i \otimes ... \otimes \sigma_n^i$, where $\sigma_k^i$ is a state of the
$k$-th subsystem and the $p_i>0$ are probabilities. Such states $\rho$ are called inseparable or entangled. One
could say that for entangled systems the whole is indeed more than the sum of its parts. Only entangled states can
exhibit quantum non-locality \cite{bell}. In this case there is no way of reproducing the predictions of quantum
physics with classical systems, unless there is instantaneous communication over arbitrary distances.

Entanglement has recently been studied extensively in the context of quantum computation and quantum communication
\cite{ent}. Quantum computation would require the creation and maintenance of highly complex entangled states of
many subsystems. This is a difficult task because of decoherence \cite{decoherence}. A system interacts with its
surroundings, which creates entanglement between system and environment, at the same time reducing the entanglement
within the system itself.

Such considerations lead to a natural physical question: how common is complex entanglement - entanglement between
many subsystems - in nature? Are there natural systems that contain substantial large-scale entanglement between
their constituents? There seem to be two conflicting intuitions in response to this question. On the one hand,
entanglement seems to be very fragile under decoherence, which is emphasized by the difficulty of building a
quantum computer. On the other hand, measure-theoretic considerations suggest that the set of inseparable states is
much larger than the set of separable ones \cite{sepset}, which seems to imply that entanglement should be
relatively common.

One might hope to gain some insight into the above question by analyzing the behaviour of complex entangled states
under decoherence, in particular by studying how fast the entanglement disappears. This is not an easy task,
because it requires practical criteria for inseparability. However, for some special sets of states such criteria
have been found \cite{duer,spinsqueezing}, which we will use in the following.

We will consider the evolution of certain multi-qubit states under the action of partially depolarizing channels,
to be defined below, acting independently on every qubit. The multi-qubit states are our models for complex
entanglement, and the depolarizing channel is our toy model for decoherence, corresponding to {\it local} and {\it
homogeneous} noise. We ask how much local depolarization is possible such that the states are still entangled, in
this way quantifying their robustness under decoherence.

Our notion of robustness is thus different from the one of \cite{vidal}, who considered global mixing of entangled
states with locally prepared noise. Admixing locally prepared noise is different from local decoherence; it
corresponds to a scenario where the local ``decohering agents'' communicate when to add noise, and when to leave
the system undisturbed.

It also differs from the notion of entanglement persistency proposed in \cite{briegel}, which is defined as the
minimum number of local measurements needed in order to definitely disentangle a given state. This definition
involves an optimization over all possible local measurements. It thus corresponds to intelligent and informed
adversaries who optimize coordinated decohering measurements in order to most efficiently destroy the entanglement
of a state. The depolarizing channels which we consider here, applied to each subsystem, correspond to independent
measurements in random bases on each local component of the entangled state.

We will consider qubits with basis states $|0\rangle$ and $|1\rangle$. The {\it completely} depolarizing channel is
defined by the transformation $|i\r \l j| \longrightarrow \delta_{ij} \frac{1}{2} \openone$, where $\openone=|0\r
\l 0|+|1\r \l 1|$. When applied to a subsystem $A$ of a composite system $AB$ this corresponds to the transformation
$\rho_{AB} \longrightarrow \frac{1}{2} \openone \otimes \mbox{Tr}_A \rho_{AB}$. All correlations between the system
and its environment are destroyed, and the system itself is put into the completely mixed state.

The {\it partially} depolarizing channel $C_d$ is defined by applying the completely depolarizing channel with a
probability $d$, and applying the identity transformation with a probability $1-d$. This corresponds to the
transformation
\be
|i\r\l j| \longrightarrow (1-d) |i\r \l j| + d \delta_{ij} \frac{1}{2} \openone.
\label{channel}
\ee
If the density matrix of the system is written in the basis of Pauli matrices, one can easily convince oneself that
(\ref{channel}) corresponds to multiplying all $\sigma_i$ by the {\it scaling factor} $s=1-d$, while the unit
matrix $\openone$ remains unaffected.

An interesting physical realization of a partially depolarizing channel is by random measurements. One can show
that if a qubit is subjected to a Von Neumann measurement in a basis chosen uniformly at random, the effect on the
density matrix corresponds to a partially depolarizing channel with $d=2/3$.

We will refer to $d$ as the amount of local depolarization. Denote the product of partial depolarizing channels
with depolarization $d$ on each qubit by $C_d^{\otimes n}$. For a given $n$-party entangled state $\rho$, we will
be interested in the critical amount of depolarization $d_{crit}$  where $C_d^{\otimes n}(\rho)$ becomes separable,
or - in the absence of a necessary and sufficient condition - ceases to fulfill certain sufficient conditions for
entanglement \cite{spinsqueezing,pereshoro}. In both cases $d_{crit}(\rho)$ will be a quantitative signature of the
robustness of the entanglement in that state.

There are two families of states where we have been able to obtain explicit results in the limit of $n$, the number
of subsystems, going to infinity, namely GHZ states and spin-squeezed states \cite{sss} of $n$ qubits. Our main
result is that, surprisingly, in both cases a very substantial amount of local depolarization does not destroy the
multiparty entanglement. We have also compared GHZ states to other generic entangled states for $n=3$ and $n=4$
qubits, with (to us) counter-intuitive results.

For the following discussion we find it convenient to define $P_0=|0\rangle \langle 0|$, $P_1=|1\rangle \langle
1|$, $\sigma_+=|0\rangle
\langle 1|$, and $\sigma_-=|1\rangle
\langle 0|$. In this basis the partially depolarizing channel (\ref{channel}) corresponds to the following transformations:
\begin{eqnarray}
P_{0,1} &\longrightarrow & \frac{1+s}{2} P_{0,1} +\frac{1-s}{2} P_{1,0} \nonumber\\ \sigma_{+,-} &\longrightarrow
&s
\sigma_{+,-}.
\label{channel1}
\end{eqnarray}
Recall that $s=1-d$.

Let us first discuss the case of GHZ states. An $n$-qubit GHZ state $1/\sqrt{2} (|00 \ldots 0\rangle + |11
\ldots 1\rangle)$ has the density matrix
\begin{equation}
\rho=\frac{1}{2}(P_0^{\otimes n}+P_1^{\otimes n}+\sigma_+^{\otimes n}+\sigma_-^{\otimes n}).
\end{equation}
Application of the channel (\ref{channel1}) to every qubit multiplies the off-diagonal terms $\sigma_+^{\otimes n}$
and $\sigma_-^{\otimes n}$ by $s^n$. The diagonal terms $P_0^{\otimes n}$ and $P_1^{\otimes n}$ give rise to new
diagonal terms of the form $P_0^{\otimes k} \otimes P_1^{\otimes (n-k)}$ and all permutations thereof, for $k$
ranging from $0$ to $n$. The coefficients $\lambda_k$ of these terms are given by
\begin{equation}
\lambda_k=\frac{1}{2}\left[ \left( \frac{1+s}{2} \right)^k \left( \frac{1-s}{2} \right)^{n-k}
+ \left( \frac{1+s}{2} \right)^{n-k} \left( \frac{1-s}{2} \right)^k   \right].
\end{equation}

In order to analyze the entanglement of states of the above form it is sufficient to study the entanglement for
bipartite cuts \cite{duer}, where some qubits constitute one subsystem and all the other qubits the other
subsystem. For each such cut one considers the partial transposition \cite{pereshoro}. As long as the partial
transpose of a state has negative eigenvalues, the state is definitely entangled.

Take partial transposition of the first $k$ qubits as an example. The diagonal terms do not change, while
$\sigma_+^{\otimes n}=(|0\rangle\langle 1|)^{\otimes n}$ goes into an off-diagonal term between the states
$|1\rangle^{\otimes k} |0\rangle^{\otimes (n-k)}$ and $|0\rangle^{\otimes k} |1\rangle^{\otimes (n-k)}$. It is then
easy to see that the state will have non-positive partial transpose for some bipartite cut, and thus definitely be
entangled, as long as
\begin{equation}
\frac{s^n}{2}>\lambda_k
\label{criterion}
\end{equation}
for some $k$. In \cite{duer} it was moreover shown that if the state has positive partial transpose with respect to
all bipartite cuts, then it is separable, so that the above condition for inseparability is both sufficient and
necessary.

The smallest eigenvalue $\lambda_k$ is the one for $k=n/2$, if $n$ is even, or $k=m$, for $n=2m+1$. Therefore we
can immediately make a statement about entanglement after local depolarization for general $n$ (consider even $n$
for simplicity). The state is definitely entangled as long as
\begin{equation}
\frac{s^n}{2}>\left(\frac{1+s}{2}\right)^{\frac{n}{2}}\left(\frac{1-s}{2}\right)^{\frac{n}{2}}.
\end{equation}
The critical value of $s$ where the entanglement disappears is given by
\begin{equation}
s_{crit}(n)=\frac{1}{\sqrt{2^{2-\frac{2}{n}} + 1}}
\end{equation}
for even $n$. It is very remarkable that $s_{crit}$ decreases with $n$. The entanglement in the GHZ state thus
becomes more robust under local depolarization when the number of subsystems is increased. In the limit $n
\rightarrow \infty$, $s_{crit}^{\infty}=\frac{1}{\sqrt{5}}=0.447$. This means that a very large GHZ state can stand
more than 55 \% depolarization ($d_{crit}^{\infty}=1-s_{crit}^{\infty}=0.553$) of every qubit before it becomes
separable.

A natural question to ask is how the robustness of the GHZ state compares to other multiparty entangled states. We
do not know of any other family of multi-qubit states for which there is a necessary and sufficient condition for
entanglement. As a first step, we have therefore chosen to apply the partial transposition criterion
\cite{pereshoro} to some characteristic states of 3 and 4 qubits. In the 4-qubit case we have studied both the 1-3
and the 2-2 bipartite cuts. These results can be directly compared to the results in the GHZ case, where the
partial transposition criterion happens to be necessary and sufficient. Our results are collected in Table
\ref{table1}. The main - again surprising - conclusion is that GHZ states are comparatively robust.

\begin{table}
\center
\begin{tabular}{|c|c|c|}
\hline $n=3$ & \multicolumn{2}{|c|}{1-2 cut} \\ \hline $G_3$ & \multicolumn{2}{|c|}{0.443} \\ \hline
$W_3$ & \multicolumn{2}{|c|}{0.425}
\\ \hline $n=4$ & 1-3 cut & 2-2 cut \\ \hline $G_4$ & 0.423 & 0.489 \\ \hline $W_4$ & 0.423 & 0.423 \\
\hline $X_4$ & 0.416 & 0.453 \\ \hline $B_4$ &
0.468 & 0.450 \\ \hline
\end{tabular}
\vskip.2cm
\caption{Critical values of local depolarization $d$ for which several generic entangled states of 3 and 4 qubits become
PPT (with positive partial transpose). For the 4-qubit states the values for both the 1-3 and the 2-2 bipartite
cuts are given. One sees that $G_4$, the 4-qubit GHZ state, stays NPT (with negative partial transpose, and thus
entangled) up to the largest value of depolarization, $d_{crit}=0.489$. For the GHZ case PPT-ness and separability
are equivalent. The states considered were the following:
$|G_3\rangle=\frac{1}{\sqrt{2}}(|000\rangle+|111\rangle)$,$|G_4\rangle=\frac{1}{\sqrt{2}}(|0000\rangle+|1111\rangle)$,
$|W_3\rangle=\frac{1}{\sqrt{3}}(|001\rangle + |010\rangle + |100\rangle)$, $|W_4\rangle=\frac{1}{2}(|0001\rangle +
|0010\rangle + |0100\rangle + |1000\rangle)$, $|X_4\rangle=\frac{1}{\sqrt{6}}(|0011\rangle + |0101\rangle
+|0110\rangle + |1001\rangle + |1010\rangle + |1100\rangle)$, and $|B_4\rangle=\frac{1}{2}(|0000\rangle +
|0011\rangle + |1100\rangle - |1111\rangle)$.}\label{table1}
\end{table}

The states considered were the following. For $n=3$, the state $|W_3\rangle=\frac{1}{\sqrt{3}}(|001\rangle +
|010\rangle + |100\rangle)$. For a symmetric state of 3 qubits, there is just one possible bipartite cut (1-2). The
state $|W_3\rangle$ becomes PPT (with positive partial transpose) across this cut for a lower value of
depolarization than the 3-qubit GHZ state, i.e., at least according to the partial transposition criterion, it is
less robust under local depolarization.

For $n=4$, we have studied three different states, namely $|W_4\rangle=\frac{1}{2}(|0001\rangle + |0010\rangle +
|0100\rangle + |1000\rangle)$, $|X_4\rangle=\frac{1}{\sqrt{6}}(|0011\rangle + |0101\rangle +|0110\rangle +
|1001\rangle + |1010\rangle + |1100\rangle)$, and $|B_4\rangle=\frac{1}{2}(|0000\rangle + |0011\rangle +
|1100\rangle - |1111\rangle)$ \cite{briegel}. There is only a single instance where one of them is more robust than
the 4-qubit GHZ state: the 1-3 cut for $|B_4\rangle$ remains NPT for higher values of depolarization. However, if
the 2-2 cut is also taken into account, again the GHZ state remains NPT for higher $d$ values.

It is worth noting that the state $|X_4\rangle$, which can also be written as
$|X_4\r=\frac{1}{\sqrt{6}}(|00\rangle|11\rangle + 2 |\psi_+\rangle |\psi_+\rangle + |11\rangle |00\rangle)$
originally has more entanglement across the 2-2 cut than the GHZ state: the Von Neumann entropy of the reduced
state is 1.252 compared to 1 for the GHZ state (for pure states, the Von Neumann entropy is a good measure of
entanglement, cf. \cite{ent}). Nevertheless it reaches the boundary of PPT states before the GHZ state. A more
trivial example for the fact that the amount of entanglement and its robustness do not have a simple relation is
provided by any number of independent shared 2-qubit singlet states. Then the collective state is exactly as robust
(or fragile) as an individual singlet, which has $d_{crit}=1-s_{crit}(2)=0.423$.

Remarkably, there is also no positive correlation between the Schmidt rank of the states with respect to the 2-2
cut (i.e. the number of terms in the corresponding Schmidt decomposition) and their robustness under local
depolarization, cf. \cite{eisert}. The state $|X_4\rangle$ has Schmidt rank 3, nevertheless it is less robust than
the GHZ state, which has Schmidt rank 2. Again, an even simpler example is provided by a state of two independent
shared singlets, $|S_4\rangle=\frac{1}{2}(|00\rangle |00\rangle +|01\rangle |01\rangle +|10\rangle |10\rangle
+|11\rangle |11\rangle)$, which has Schmidt rank 4, but is less robust than the GHZ state of 4 qubits.

There is one more class of states for which we were able to study the entanglement robustness for a general number
of qubits $n$, in particular $n$ tending towards infinity, namely states exhibiting spin squeezing \cite{sss}. An
$n$-qubit state $|\psi\rangle$ is called spin squeezed, if it satisfies
\begin{equation}
\xi^2=\frac{n\langle J_x^2 \rangle}{\langle J_y \rangle^2 + \langle J_z \rangle^2} < 1
\label{sscrit}
\end{equation}
for three appropriately chosen orthogonal directions $x$, $y$, and $z$. Here $J_{x,y,z}=\frac{1}{2}\sum
\limits_{i=1}^{n} \sigma_{x,y,z}^{(i)}$ are the total angular momentum operators, and $\langle J_x^2 \rangle=
\langle \psi| J_x^2 |\psi\rangle$ etc. It was shown in \cite{spinsqueezing} that spin squeezing is a sufficient
condition for entanglement. Such states can e.g. be generated from an initial product state of the $n$ qubits
through time evolution with a Hamiltonian of the general form $H=\chi J_x^2$.

For sufficiently large $n$, there are states for which $\xi^2$ is very small, $\langle J_z \rangle=\zeta
\frac{n}{2}$ with $\zeta$ close to 1, and $\langle J_x \rangle=\langle J_y \rangle=0$ \cite{sss}.
The local depolarizing channels (\ref{channel}) can be quite easily applied in the present situation. If the state
is written in the basis of Pauli matrices, then simply all $\sigma_i$ are multiplied by a factor $s=1-d$. One finds
\begin{eqnarray}
\langle J_z \rangle &\rightarrow & s \langle J_z \rangle \nonumber\\
\langle J_x^2 \rangle=\frac{n}{4}+\frac{1}{4}\sum \limits_{i \neq j} \langle \sigma_x^{(i)}\sigma_x^{(j)}\rangle
&\rightarrow &\frac{n}{4}+\frac{s^2}{4}\sum \limits_{i \neq j} \langle
\sigma_x^{(i)}\sigma_x^{(j)}\rangle = \nonumber\\
&&=(1-s^2)\frac{n}{4}+ s^2 \langle J_x^2 \rangle,
\end{eqnarray}
where again $s=1-d$, and thus for the squeezing parameter $\xi^2_s$ after depolarization:
\begin{equation}
\xi^2_s=\frac{1-s^2}{\zeta^2 s^2}+\xi^2_0,
\end{equation}
where $\xi^2_0$ is the original squeezing parameter.

This means that under local depolarization the states definitely remain entangled for all scaling factors larger
than
\begin{equation}
s_{crit}=\frac{1}{\sqrt{1+\zeta^2(1-\xi_0^2)}}.
\end{equation}
This is equal to $\frac{1}{\sqrt{2}}=0.707$ in the limit of $\zeta=1$ and $\xi_0^2=0$, which is approached for $n
\rightarrow \infty$. This means that in this limit the spin-squeezing entanglement can stand more than 29 \% of
local depolarization. Note that (\ref{sscrit}) is a sufficient, but not a necessary, condition for entanglement, so
the states may well be entangled for even larger values of depolarization.

In the present work we have used local partially depolarizing channels as our model for decoherence, corresponding
to local and homogeneous noise. Our results emphasize that the robustness of multi-party entangled states will in
general depend on the nature of the decoherence process. In particular, GHZ states have turned out to be
surprisingly robust under local depolarization. On the other hand, GHZ entanglement is of course fragile under some
other operations, such as the loss of a single qubit, or the measurement of a single qubit in the
$\{|0\rangle,|1\rangle\}$ basis. For the latter scenario it is however worth noting that the measurement has to be
performed {\it with certainty} in order to destroy the entanglement. To show this consider the state generated from
the $n$-qubit GHZ state $|G_n\rangle$ by performing local measurements in the $\{|0\rangle,|1\rangle\}$ basis on
every qubit with a probability $p$. {\em Certain} measurement corresponds to $p=1$. This state is given by
\begin{equation}
\frac{1-(1-p)^n}{2}(P_0^{\otimes n}+P_1^{\otimes n})+(1-p)^n |G_n\rangle \langle G_n|.
\end{equation}
It is entangled for all $p<1$, as can be seen from its partial transpose. Of course, the coefficient of the
inseparable term decreases exponentially.

One should certainly be careful in drawing general conclusions from a single simple model. Nevertheless, our
results show that multi-party entanglement can be surprisingly robust under decoherence, and that the robustness
can even increase with the number of parties. This suggests that large-scale entanglement could be more frequent in
natural systems than one might have expected. Persistent entanglement would definitely require the presence of an
entanglement-generating process - interactions between the subsystems, possibly in combination with external
excitation - that competes with the decoherence. We believe that the competition between entanglement generation
and decoherence in multi-party systems is a promising and fascinating topic for future research.

C.S. would like to thank H. Briegel, L. Hardy, and L. Henderson for a very useful discussion. C.S. has been
supported by the QIPC program of the European Union (project QuComm) and the Austrian Science Foundation (FWF,
project S6503). J.K.'s effort is sponsored by the Defense Advanced Research Projects Agency (DARPA) and Air Force
Laboratory, Air Force Materiel Command, USAF, under agreement number F30602-01-2-0524. The U.S. Government is
authorized to reproduce and distribute reprints for Governmental purposes notwithstanding any copyright annotation
thereon. J.K. also thanks the Erwin-Schr\"odinger Institute, Vienna, where part of this research was conducted, for
support and hospitality. Both authors are grateful for the hospitality of the Benasque Center for Science.

\end{document}